\documentclass[journal,twoside,web]{ieeecolor}
\usepackage{IEEE_generic}
\usepackage{cite}
\usepackage{amsmath,amssymb,amsfonts}
\usepackage{algorithmic}
\usepackage{graphicx}
\usepackage{float}
\usepackage{adjustbox}
\usepackage{caption}
\usepackage[noend,ruled,vlined]{algorithm2e}
\usepackage{stfloats}
\usepackage[justification=justified]{caption}
\usepackage{textcomp}
\usepackage[utf8]{inputenc}
\usepackage{longtable}
\usepackage{url}
\usepackage{xr}
\usepackage{multirow}
\usepackage{booktabs}
\renewcommand{\arraystretch}{1.4}
\usepackage[switch,columnwise]{lineno}
\usepackage{pdfpages}
\usepackage{amsmath,amssymb,amsfonts}
\usepackage{float}
\usepackage{adjustbox}
\usepackage{stfloats}
\usepackage{textcomp}
\usepackage{longtable}
\usepackage{url}
\usepackage{xr}
\usepackage{multirow}
\usepackage[colorlinks]{hyperref}
\usepackage{nicefrac}       % compact symbols for 1/2, etc.
\usepackage{microtype}      % microtypography
\usepackage{booktabs}
\usepackage{multirow}
\usepackage{siunitx}
\usepackage{graphicx}
\definecolor{Blue}{rgb}{0.0, 0.00, 1.00}
\usepackage{array}
\hypersetup{
  colorlinks,
  citecolor=teal,
  linkcolor=Blue,
  urlcolor=Blue}
\usepackage[font=footnotesize,justification=justified,belowskip=2pt,aboveskip=2pt]{caption}

\DeclareUnicodeCharacter{2081}{\ensuremath{{}_1}}

\definecolor{bondiblue}{rgb}{0.0, 0.58, 0.71}
\definecolor{brightcerulean}{rgb}{0.11, 0.62, 0.74}

\usepackage{chngcntr}
\counterwithin*{subsubsection}{subsection}
\renewcommand{\thesubsubsection}{\thesubsection.\alph{subsubsection}}

\def\BibTeX{{\rm B\kern-.05em{\sc i\kern-.025em b}\kern-.08em
    T\kern-.1667em\lower.7ex\hbox{E}\kern-.125emX}}
\begin{document}
\title{HydraViT: Adaptive Multi-Branch Transformer \\ for Multi-Label Disease Classification \\ from Chest X-ray Images}
\author{\c{S}aban \"{O}zt\"{u}rk, M. Yi\u{g}it Tural{\i}, and Tolga \c{C}ukur$^*$
\thanks{This research is funded by the Scientific and Technological Research Council of Turkey (TUB\.{I}TAK) under grant number 118C543. \c{S}. \"{O}zt\"{u}rk, M. Y. Tural{\i}, and T. Çukur are with the Department of Electrical and Electronics Engineering, and the National Magnetic Resonance Research Center, Bilkent University, Ankara, Turkey (e-mails: saban.ozturk@amasya.edu.tr, yigit.turali@ug.bilkent.edu.tr, cukur@ee.bilkent.edu.tr). Ş. Öztürk is also with the Amasya University, Amasya, Turkey.}
}

\maketitle

\begin{abstract}
Chest X-ray is an essential diagnostic tool in the identification of chest diseases given its high sensitivity to pathological abnormalities in the lungs. However, image-driven diagnosis is still challenging due to heterogeneity in size and location of pathology, as well as visual similarities and co-occurrence of separate pathology. Since disease-related regions often occupy a relatively small portion of diagnostic images, classification models based on traditional convolutional neural networks (CNNs) are adversely affected given their locality bias. While CNNs were previously augmented with attention maps or spatial masks to guide focus on potentially critical regions, learning localization guidance under heterogeneity in the spatial distribution of pathology is challenging. To improve multi-label classification performance, here we propose a novel method, HydraViT, that synergistically combines a transformer backbone with a multi-branch output module with learned weighting. The transformer backbone enhances sensitivity to long-range context in X-ray images, while using the self-attention mechanism to adaptively focus on task-critical regions. The multi-branch output module dedicates an independent branch to each disease label to attain robust learning across separate disease classes, along with an aggregated branch across labels to maintain sensitivity to co-occurrence relationships among pathology. Experiments demonstrate that, on average, HydraViT outperforms competing attention-guided methods by 1.2\%, region-guided methods by 1.4\%, and semantic-guided methods by 1.0\% in multi-label classification performance.

\end{abstract}

\begin{IEEEkeywords}
multi-label classification, Chest X-ray, deep learning, transformer, label co-occurrence. 
\end{IEEEkeywords}

\bstctlcite{IEEEexample:BSTcontrol}

\section{Introduction}

%motivasyon ve konuya genel giriş
The prevalence of thoracic diseases is a growing concern that poses a significant threat to human health. Lung cancer, the second most common cancer globally, accounts for 11-12\% of all cancer cases and is responsible for approximately 18\% of cancer-related deaths \cite{SAHA2022e493}. Aspiration pneumonia, another aggressive thoracic disease, is responsible for about 2-3\% of all deaths in developed countries \cite{gupte2022mortality}. A prominent imaging technology for early diagnosis of these deadly conditions is chest X-ray (CXR), which is cost-efficient compared to other common modalities. However, the ever-increasing number of CXR scans, complex pathologies, variable lesion sizes, and subtle texture changes can compromise the accuracy of radiological readings. These challenges are further exacerbated by operator biases in developing countries with relatively limited accumulation of radiological expertise \cite{kruk2018mortality}. Therefore, the development of computer-aided diagnosis (CAD) algorithms that can automatically diagnose thoracic diseases from CXR scans can serve to improve efficiency and accuracy in radiological assessments.

%Mevcut literatür biraz ve problemler
The mainstream CAD approach for diagnosing thoracic diseases rests on the extraction of CXR features to help identify and locate pathological regions, followed by classification based on the extracted features to identify the presence of one or more diseases \cite{GUAN202038, WANG2021101846,su_concurency}. Convolutional neural networks (CNNs) have arguably become the de facto standard in the extraction of CXR features, given their efficiency in learning visual features for downstream imaging tasks \cite{kim2022transfer,celard2023survey}. Yet, CNN models use compact local filters with static weights for feature extraction, and they are typically trained with conventional softmax output layers that compromise between sensitivity to individual classes versus co-occurrence relationships among classes. In turn, CNN can elicit suboptimal performance for identification of complex pathology in thoracic diseases. Primary causes of performance loss include the following: (1) The size, location and appearance of pathology in CXR images show high variability across disease classes and across subjects, compromising generalization abilities; (2) Separate pathology or multiple instances of a given pathology can frequently co-occur in CXR images of anatomy, and the co-occurrence statistics show high heterogeneity across pathology labels, making it difficult to maintain consistent performance across labels \cite{9336317,WANG2023104488}.

A first group of CXR studies have proposed to augment CNN models with attention modules or attention-based masks in order to improve network focus on small-sized pathology  \cite{ZHU2022102137,JUNG202334,GUAN202038,chen2019lesion,CHEN2020221,hossain2022novel}. Despite the efficiency and promising performance of attention-augmented CNNs, they can still show limited capture of the long-range context under multiple distributed or large-sized lesions, and limited generalization performance across subjects. A second group of studies have instead proposed vision transformer (ViT) models based on self-attention mechanisms to improve capture of long-range context and to improve generalization at the expense of elevated computational load \cite{LEE2022143,li2022modeling}. To maintain a desirable trade-off between performance and efficiency, hybrid models that integrate CNN and ViT blocks have also been proposed \cite{WANG2021101846,wang2019thorax,sriker2022class,chen2022thorax,ZHU2022102137,teixeira2020dualanet,jung2022graph}. While these deep learning models have been adopted to learn representative features in multi-label CXR studies, they often neglect to-occurrence relationships between separate pathology \cite{LEE2022143,JUNG202334,WANG2020354,KABIR2023118942}. Pre-defined hierarchical relationships between pathology labels have been considered previously to construct multi-label CXR classifiers \cite{CHEN2020101811,PHAM2021186}. However, these previous methods are typically trained to optimize the prediction accuracy for an aggregate output vector across classes, yielding heterogeneous classification performance across individual pathology labels.

%bizim teknik nedir,problemleri nasıl çözdük
Here, we introduce a novel adaptive multi-branch transformer model for multi-label disease classification from CXR images, named HydraViT. Our proposed model leverages a hybrid architecture composed of a convolutional spatial encoder module to efficiently extract feature maps of CXR images, and a transformer-based context encoder module to capture long-range contextual relationships across image patches and co-occurring pathology. To avoid bias due to co-occurrence patterns among disease labels, multi-task learning is performed based on a multi-branch output layer as inspired by recent machine learning studies \cite{mullapudi2018hydranets,velasco2022hydranet}. Yet, differently from previous studies on multi-task learning, we propose a novel loss function that adaptively weights each output branch to further improve the learning of pathology co-occurrences. To our knowledge, HydraViT is the first model in the literature that adopts multi-task learning for each pathology label in multi-label CXR classification. 
Our main contributions are summarized below:
\begin{itemize}
\item HydraViT is a novel hybrid convolutional-transformer model that performs multi-tasking to improve reliability in multi-label disease classification from CXR images. 

\item HydraViT uses a transformer-based context encoder to capture long-range context and co-occurrence relationships between distinct pathology. 

\item HydraViT uses distance-based adaptive weights to account for variable co-occurrence statistics between separate pathology and thereby improve the homogeneity of model performance across classes. 
\end{itemize}

\section{Related Work}

\subsection{Deep Learning for CXR Classification}
The traditional framework for automated analysis of CXR images rests on the use of hand-constructed features and manual operator intervention, which undermine classification performance \cite{kumar2014distinguishing,alfadhli2017classification}. With the advent of deep learning, performance leaps have been attained in CXR analysis based on a variety of different CNN architectures \cite{rajpurkar2017chexnet,rubin2018large,baltruschat2019comparison}. Several lines of improvements in CNN models to have been considered for including deeper architectures \cite{WANG2020354}, integration of recurrence dependencies \cite{yao2017learning}, pyramidal architectures \cite{li2022modeling}, and ensembling to capture a diverse array of features \cite{rubin2018large,chen2019dualchexnet,teixeira2020dualanet,CHEN2020221,yang2023performance,hermoza2020region}. Among recent studies on CNN-based CXR classification, \cite{KABIR2023118942} uses deep feature selection to extract most informative image features. \cite{9336317} uses feature selector and integrator branches to learn discriminative features of pathology. \cite{9430552,8961143,jung2022graph} use graph features to capture semantic similarities between image features. Several recent studies have employed diffusion-based methods \cite{dar2022adaptive,ozbey2022unsupervised} to improve feature reliability based on stable diffusion \cite{chambon2022adapting} and latent diffusion models \cite{packhauser2022generation}.

A primary limitation of conventional CNN models concerns the use of static, local filter weights that can compromise generalization to atypical anatomy that varies in size, location, shape across subjects. Recent CXR classification studies have considered to use attention mechanisms either as augmentation to CNN backbones or as self-attention in ViT backbones to to improve generalization and to guide the focus of the model towards disease-relevant regions \cite{chen2019lesion,GUAN202038,hossain2022novel,JUNG202334}. Multiple attention mechanisms across different dimensions have been deployed including combination of channel, element, scale attention \cite{WANG2021101846,yan2018weakly}, channel and spatial attention \cite{9336317,LEE2022143}, class and label attention \cite{sriker2022class,su_concurency}, and multi-head self-attention \cite{li2022modeling}. Among recent studies, \cite{ZHU2022102137} proposes PCAN that uses a pixel-wise attention branch. \cite{wang2019thorax} proposes Thorax-Net with an attention branch to exploit the correlation between class labels and locations of pathology. DuaLAnet by \cite{teixeira2020dualanet} includes two asymmetric attention networks to extract more discriminative features. \cite{chen2022thorax} introduces PCSANet that uses a shuffle attention module to prioritize features related to pathology. Although these previous methods have focused on architectural improvements to CXR classification models so as to extract task-relevant features by focusing on pathology, they can elicit heterogeneous classification performance when multiple co-occurring pathologies are present in multi-label classification tasks. To address this limitation, HydraViT uniquely uses multi-task learning with a separate output branch for each label and adaptively weights each branch to cope with variable co-occurrence statistics across labels.

\begin{figure*}[t!]   
  \centering
\includegraphics[width=1.0\linewidth]{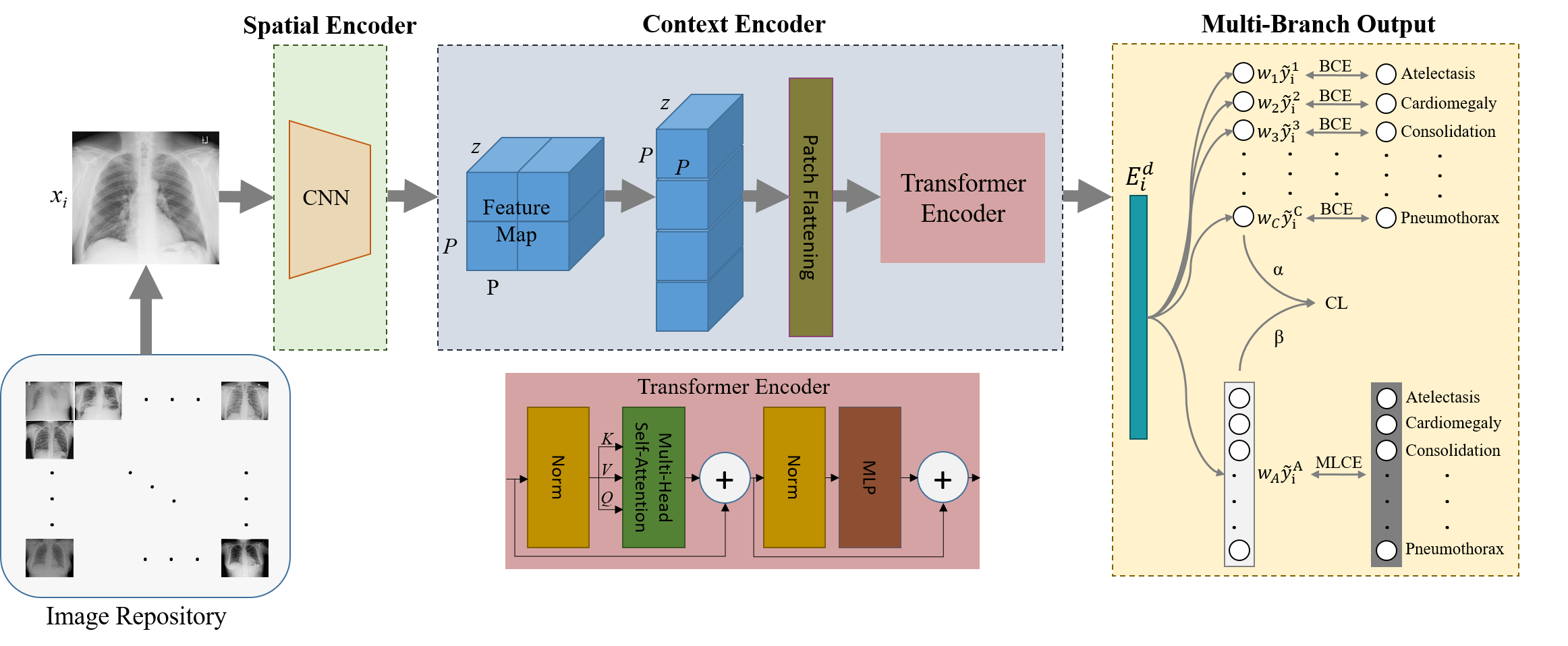}  
\caption{HydraViT consists of a CNN-based spatial encoder module, a transformer-based context encoder module, and a multi-branch output module to maintain sensitivity to individual labels and to capture label co-occurrence relationships in multi-label CXR classification.}
  \label{fig:Fig1}
\end{figure*}

\subsection{Multi-Label Classification}
A common approach for multi-label CXR classification is based on model training via cross-entropy loss on a multi-dimensional output vector that spans over all examined classes produced by a softmax layer \cite{CHEN2020221,WANG2021101846,WANG2020354,rajpurkar2017chexnet,https://doi.org/10.1002/ima.22773}. However, this conventional approach can compromise sensitivity to individual labels and can elicit suboptimal capture of co-occurrence relationships between separate labels \cite{teixeira2020dualanet,chen2019lesion}. To improve the learning of co-occurrence statistics between labels, a group of studies propose to refine label predictions via add-on network modules such as a co-occurrence module \cite{chen2019multi}, a graph module \cite{aviles2019graphx}, a recurrent module \cite{yao2017learning}, or a spatial-and-channel encoding module \cite{9336317} to capture the semantic dependencies between separate labels. As an alternative approach, other studies propose to use modified loss functions such as weighted cross-entropy \cite{zhou2018weakly} or multi-label softmax losses \cite{ge2018chest} to leverage correlations among labels for improved classification performance. Commonly, these previous studies help emphasize label co-occurrence relationships during class predictions. However, since they utilize a single multi-dimensional output vector, they can be suboptimal in preserving sensitivity to individual labels. A recent study aimed to address this issue by an ensemble CNN model, where a separate CNN predicted the label for each class \cite{lopez2021automated}. While this improves sensitivity to individual labels, it ignores label co-occurrence relationships. 

To attain sensitivity to both individual labels and their co-occurrence relationships, HydraViT employs a multi-branch output module that maps contextual embeddings extracted by its transformer-based encoder onto class labels. Unlike previous multi-label methods, HydraViT simultaneously uses segregated uni-dimensional output variables for each individual label and an aggregated multi-dimensional output vector across labels. Adaptive weights are assigned to each output variable prior to the calculation of cross-entropy losses, augmented with a consistency loss between the individual and aggregated outputs to maintain consistency between their predicted pathology labels. While several imaging and computer vision studies have considered the use of separate network branches for each individual label in classification models \cite{mullapudi2018hydranets,lopez2021automated,velasco2022hydranet}, no previous study has proposed concurrent use of individual and aggregated branches whose predictions are aligned with a consistency loss to our knowledge.

\section{Theory}

\subsection{Problem Definition}
Let us assume a training set of CXR images and corresponding disease labels $\left\{ x_{i},y_{i} \right\}_{i=1}^{N}$, where $N$ is the number of training samples. $x_{i}\in\mathbb{R}_{}^{H,W}$ is $i$th image with ($H$,$W$) denoting the image size across spatial dimensions. $y_{i}\in\mathbb{Z}_{2}^{C}$ is the $C$-dim label vector where $C$ is the number of disease classes, and $y_{i}^{c}\in \left\{ 0,1 \right\}$ serves as an indicator for the $c$th class (0: absent, 1: present). To learn the required mapping for multi-label classification, i.e., $f:x_i \to y_{i}$, a mainstream approach employs cross-entropy loss \cite{JUNG202334,WANG2021101846,CHEN2020221,zhou2018weakly,chen2019lesion,8961143,wang2019thorax}. Yet, conventional cross-entropy loss reflects an aggregate measure across all disease labels, so it does not explicitly consider the co-occurrence relationships among distinct pathology. In turn, a simple adoption of cross-entropy loss in multi-label classification can result in suboptimal performance.

\subsection{HydraViT}
To address the above-mentioned problems, HydraViT leverages multi-task learning based on a hybrid architecture where a CNN-based spatial encoder extracts lower-dimensional maps of local features followed by a transformer-based context encoder captures contextualized embeddings within and across pathologies in the input CXR image (Figure \ref{fig:Fig1}). Multi-task learning is then exercised via a synergistic combination of dedicated output variables for each individual label and a multi-dimensional output vector aggregated across labels. To maintain sensitivity to both individual labels and label co-occurrence relationships, the learnable weighting of output variables is used in conjunction with a consistency loss between the individual and aggregated output variables. Network components and learning procedures for HydraViT are described below. 

\subsubsection{Spatial Encoder (SE)} The CNN-based SE module with parameters $\theta_{SE}$ is used to extract local spatial features of CXR images and lower dimensionality of feature maps prior to context encoding. Given the input image $x_{i}$, a low-dimensional latent representation $m_{i} \in \mathbb{R}_{}^{H,W,z}$ is derived as $f_{SE}:x_{i}\to m_{i}$, where $z$ is the dimensionality of feature channels:
\begin{align}
    m_{i}= Pool \left( \sigma \left( Conv \left( \cdots Pool\left( \sigma \left( Conv \left( x_i \right) \right)\right) \cdots \right) \right) \right)
    \label{eq:Eq1_0}
\end{align}
where $Pool$ denotes a maximum pooling layer across a 2$\times$2 neighborhood for two-fold downsampling, $\sigma$ is an ReLU activation function, $Conv$ denotes a convolutional block. 

\subsubsection{Context Encoder (CE)} The transformer-based CE module with parameters $\theta_{CE}$ projects spatially-encoded feature maps onto contextualized embedding vectors, $f_{CE}:m_{i} \to E_{i}$, to capture long-range spatial relationships both within and across individual pathologies. For this purpose, $m_{i}$ from the SE module is split into $N_{p} =r_{}^{2} /P_{}^{2}$ non-overlapping patches of size ($P$, $P$) with $P=r/2$, and flattened to $zP_{}^{2}$-dimensional vectors. The transformer encoder first projects the flattened patches onto an $N_{D}$-dimensional space through learnable linear projections and positional encodings:
\begin{align}
    E_{i}^{0}=\left[ \left( m_{i} \right)_{}^{1}P_{E}^{};\left( m_{i} \right)_{}^{2}P_{E}^{};\cdots;\left( m_{i}\right)_{}^{N_{P}}P_{E}^{}\right]+P_{E}^{pos}
    \label{eq:Eq1}
\end{align}
where $E_i^{0}\in \mathbb{R}_{}^{N_{p},N_{D}}$ denote patch embeddings, $(m_{i})^{p} \in \mathbb{R}_{}^{P_{}^{2}}$ denotes the $p$th patch, $P_{E}$ and $P_{E}^{pos}$ are the linear projections and positional encodings, respectively. Next, path embeddings are processed via $L$ transformer blocks, each comprising a cascade of layer normalization ($Norm$), multi-head self-attention ($MHSA$), and multi-layer perceptron ($MLP$) layers \cite{9758823}. The $l$th block performs the following computations:
\begin{align}
    \bar{E}_{i}^{l}&=MHSA\left( Norm\left( E_{i}^{l-1} \right) \right)+E_{i}^{l-1} \\
    E_{i}^{l}&=MLP\left( Norm\left( \bar{E}_{i}^{l} \right) \right)+\bar{E}_{i}^{l}
    \label{eq:Eq3}
\end{align}
The output of the CE module $E_i^{L}$ is taken as the contextualized embedding vector $E_{i}$. Please note that the conventional output head in a transformer encoder would map the embedding vector onto activations $o_i^{1,...,C}$ in $C$ output neurons, and leverage a softmax function ($\varsigma$) to compute probabilities for separate classes: $\varsigma( {o_i^{j}} )=\frac{e_{}^{o_i^{j}}}{\sum_{k=1}^{C}e_{}^{o_i^{k}}},\ \text{for}\ j=1,...,C$. Because this formulation enforces a single class to dominate over the remaining classes in the output vector, it can be ineffective in capturing co-occurrence relationships between separate labels.

\subsubsection{Multi-Branch Output (MBO)} Given $E_{i}$, the MBO module computes $C$ uni-dimensional output variables for each label $\tilde{y}_i^{1},\tilde{y}_i^{2},...,\tilde{y}_i^{C}$ such that $\{ \tilde{y}_i^c \in \mathbb{R}_{}^{1}: 0 \le \tilde{y}_i^c \le 1\}$, along with a multi-dimensional output vector aggregated across labels $\tilde{y}_{i}^{A} \in \mathbb{R}_{}^{C}$ such that $\{ [\tilde{y}_i^A]^c \in \mathbb{R}_{}^{1}: 0 \le [\tilde{y}_i^A]^c \le 1\}$.  The resultant mapping is given as $f_{MHO}:E_{i}^{d}\to [\tilde{y}_i^{1}],[\tilde{y}_i^{2}],...,[\tilde{y}_i^{C}],[\tilde{y}_{i}^{A}]$. A learnable weight is employed for each output variable $w_{1},w_{2},...,w_{C},w_A \in \mathbb{R}_{}^{1}$ to account for label co-occurence. These weights are initialized based on the observed ratios of samples in the training set (i.e., $w_{c}=N/\left( C*N_{c} \right)$ for the $c$th class with $N_c$ training samples; $w_A = 1/(C+1)$). The weights are incorporated in the loss terms for model training:
\begin{multline}
    \text{Loss}\left( x_{i},y_{i} \right)=\frac{1}{C}\sum_{c=1}^{C}\text{BCE}\left( y_{i}^{c}, w_c \tilde{y}_{i}^{c} \right) + \text{MLCE}\left( y_{i},w_A \tilde{y}_{i}^A \right) \\ + \text{CL} \left( \alpha\left(  [w_1\tilde{y}_i^1, ... ,w_C\tilde{y}_i^C]  \right),\beta\left(  w_A \tilde{y}_i^A \right) \right)
    \label{eq:Eq6}
\end{multline}
The first term in Eq. \ref{eq:Eq6} is binary cross-entropy loss expressed for each uni-dimensional output variable:
\begin{align}
    \text{BCE}\left( y,\tilde{y} \right)= -\left[ y\text{log}\left( \tilde{y} \right) + \left( 1-y \right)\text{log}\left( 1- \tilde{y} \right) \right]
    \label{eq:Eq4}
\end{align}
where $y$ and $\tilde{y}$ denote binary scalars that reflect the true and predicted labels for a given class. The second term in Eq. \ref{eq:Eq6} is multi-label cross-entropy loss expressed for the multi-dimensional output vector as:
\begin{align}
    \text{MLCE}\left( y^A,\tilde{y}^A \right)=\frac{1}{C}\sum_{c=1}^{C}\left( \text{BCE}\left( [y^A]^c,[\tilde{y}^A]^c \right) \right)
    \label{eq:Eq5}
\end{align}
where $y^A$ and $\tilde{y}^A$ denote binary vectors that reflect the true and predicted labels across $C$ classes. The final term in Eq. \ref{eq:Eq6} is consistency loss that enforces the predictions from the individual and aggregated output variables to be consistent with each other:
\begin{equation}
\text{CL}\left( \tilde{y}^{A,1}, \tilde{y}^{A,2}\right)=\left\| \tilde{y}^{A,1} - \tilde{y}^{A,2} \right\|_{2}
\end{equation}
where $\tilde{y}^{A,1}$ is formed by concatenating individual output variables across the label dimension and scaling the resultant vector by a factor of $\alpha$, and $\tilde{y}^{A,2}$ is derived from the aggregated output vector via scaling by a factor of $\beta$. The scaling factors are taken as learnable parameters. The training procedures for HydraViT based on the loss given in Eq. \ref{eq:Eq6} are described in Alg. \ref{alg:one}. During inference on a test CXR image, $x_{qi}$, the predictions from the individual output variables, i.e., $ [w_1\tilde{y}_{qi}^1, ... ,w_C\tilde{y}_{qi}^C]$, are used to generate the class predictions.

\SetKwInput{KwData}{Input}
\SetKwInput{KwResult}{Output}
\RestyleAlgo{ruled}
%\SetAlgoNoLine

\SetKwInOut{Parameters}{Parameters}
\begin{algorithm}[htb]
    \small
    \caption{Training procedure for HydraViT}
    \label{alg:one}
    \KwData{Dataset:$\left\{ x_{i},y_{i} \right\}_{i=1}^{N}$, $x_{i}$: CXR image, $y_{i}$: label\\
    \Indp\Indp $f_{SE}$: Spatial encoder with param. $\theta_{SE}$\\    
    $f_{CE}$: Context encoder with param. $\theta_{CE}$\\
    $\tilde{y}_{i}^{1,...,C}$: Individual output variables\\
    $\tilde{y}_{i}^{A}$: Aggregated output vector\\
    $Opt$(): Optimizer for computing param. updates
    }
    \KwResult{$\psi$: \{$w_{1}$, ..., $w_{C}$, $w_{A}$, $\alpha$, $\beta$, $\theta_{SE,CE}$\}}
     Initialize parameters.\\
     \For{$i$=$1$:$N$}{
      Compute $m_{i}$, $f_{SE}(\theta_{SE}):x_{i}\to m_{i}$\\ 
      Compute $E_{i}$, $f_{CE}(\theta_{CE}):m_{i}\to E_{i}$\\
      Project onto individual branches, $[\tilde{y}_i^{1}],[\tilde{y}_i^{2}],...,[\tilde{y}_i^{C}]$\\
      Project onto an aggregate branch, $\tilde{y}_{i}^{A} $\\
      Compute BCE for individual branches via Eq. \ref{eq:Eq4}\\
      Compute MLCE for the aggregate branch via Eq. \ref{eq:Eq5}\\
      Compute Loss based on Eq. \ref{eq:Eq6}\\
      Update model parameters: $\psi\gets \psi-Opt\left( \triangledown_{\psi} \text{Loss} \right)$\\
     }
     \Return $\psi$
\end{algorithm}

\begin{figure*}[t]
\centering
\includegraphics[width=0.675\linewidth]{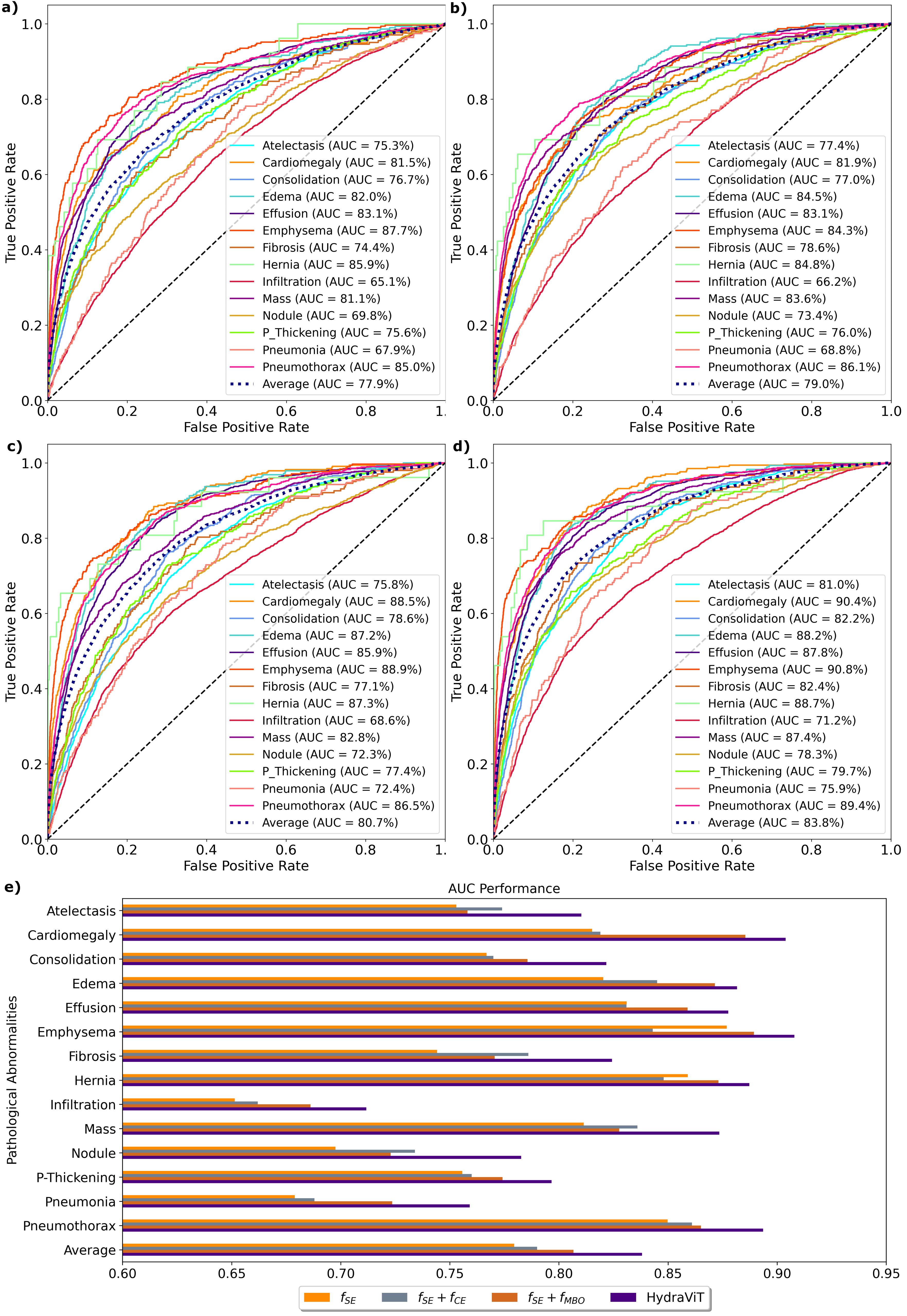}
\caption{Multi-label CXR classification performance of HydraViT and variant models. Results are shown for a variant model that contained only the SE module $f_{SE}$ that was augmented with a softmax output layer for multi-label classification, a variant model that contained the SE and CE modules $f_{SE}+f_{CE}$ augmented with a softmax output layer, and a variant model that contained the SE and MBO modules $f_{SE}+f_{MBO}$. Class-wise and class-average ROC curves for (a) $f_{SE}$, (b) $f_{SE}+f_{CE}$, (c) $f_{SE}+f_{MBO}$, and (d) HydraViT; and (e) class-wise and class-average AUC metrics for all models.}
\label{fig:fig2}
\end{figure*}

\section{Methods}
\subsection{Dataset}
Demonstrations were performed on the ChestX-ray14 dataset \cite{8099852} with 112,120 frontal-view images from 30,805 unique patients with ages 1-95 years. Of these patients, 56.49\% are male, and 43.51\% are female. The dataset includes labels for 15 classes for each CXR image, including the 'No Finding' class for healthy individuals, and 14 different pathologies (atelectasis, cardiomegaly, effusion, infiltration, mass, nodule, pneumonia, pneumothorax, consolidation, edema, emphysema, fibrosis, pleural thickening, and hernia). Note that the average size of pathological regions is approximately 7.5\% of the image size. The classes except for 'No Finding' can be simultaneously present in a given patient, yielding a multi-label classification problem. The 'No Finding' class accounts for 53.83\% of the total dataset with 60,412 samples. Major pathological abnormalities such as 'Infiltration' and 'Effusion' have sample sizes of 19,894 and 13,317, respectively, while minor pathological abnormalities such as 'Hernia' and 'Pneumonia' have sample sizes of 227 and 1,431, respectively.

Prior to modeling, all CXR images were spatially downsampled to a 224$\times$224 grid for computational efficiency. Data were split into training and test sets without any patient-level overlap, while preserving the ratios between the number of samples for separate classes. The training split contained 86,524 images, whereas the test split contained 25,596 images.

\begin{figure*}[t]
\centering
\includegraphics[width=0.95\linewidth]{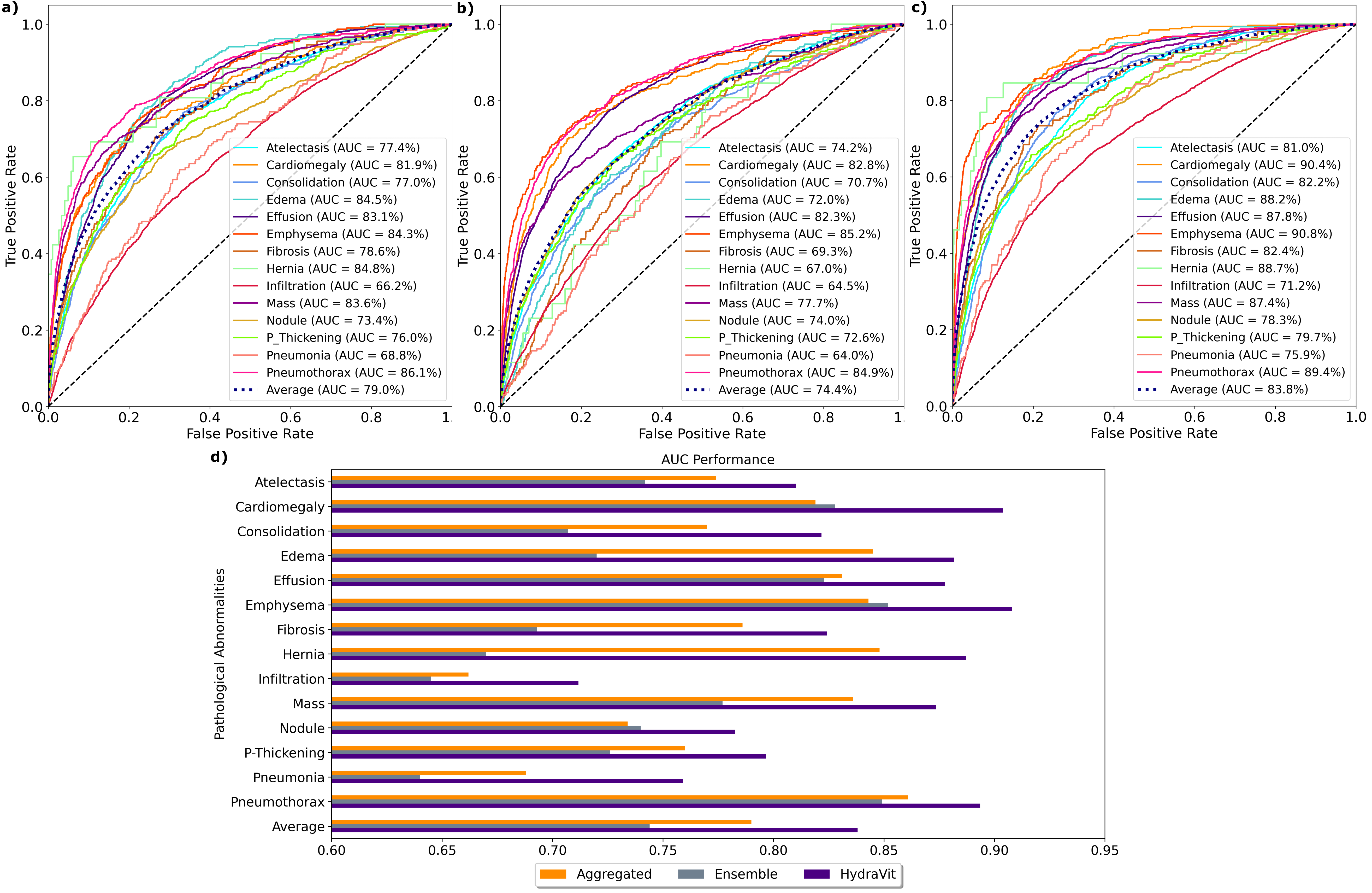}
\caption{Multi-label CXR classification performance of HydraViT and variant models. Results are shown for a variant model that trained a single $f_{SE}+f_{CE}$ architecture with an aggregated output vector across labels, and a variant model that ensembles separate $f_{SE}+f_{CE}$ architectures trained for each individual label. Class-wise and class-average ROC curves for (a) the aggregated variant, (b) the ensemble variant, and (c) HydraViT; and (d) class-wise and class-average AUC metrics for all models.}
\label{fig:fig3}
\end{figure*}

\subsection{Implementation Details}
HydraViT leverages a hybrid network architecture with a spatial encoder module, a context encoder module and a multi-branch output layer. The spatial encoder module was implemented based on a pre-trained VGG16 architecture with $z$=512 and $r$=7. This architecture consists of 13 convolutional layers with a 3$\times$3 kernel size, ReLU layers, 5 max-pooling layers with a 2$\times$2 kernel size, and 3 fully connected layers (FCL). The context encoder module used 12 transformer blocks with a projection dimension of 512, 20 attention heads, $P$=4 resulting in $N_{P}$=4, and $d$=25,088. The $w_{A}$ value in the multi-head output block was initialized as $1/(C+1)$. The $\alpha$ and $\beta$ parameters were initialized randomly in the range of [0 5]. HydraViT was implemented using the TensorFlow framework and executed on an NVidia RTX 3090 GPU. Models were trained via the Adam algorithm with a batch size of 35, a learning rate of $10_{}^{-4}$, and 120 epochs.

\subsection{Competing Methods}
HydraViT was compared against several state-of-the-art deep-learning models for multi-label CXR classification. Three main groups of competing methods were considered: attention-guided, region-guided, and semantic-guided methods.

\subsubsection{Attention-guided methods} 
\textbf{\textit{PCAN}}: Pixel-wise classification and attention network (PCAN) \cite{ZHU2022102137} extracts mid-level CXR image features via a CNN, and uses pixel-wise branches for classification.

\textit{$\textbf{A}_{}^{3}$}\textbf{\textit{Net}}: Triple-attention learning ($\text{A}_{}^{3}$Net) \cite{WANG2021101846} extracts features via DenseNet121, and uses channel, element, scale attention.

\textbf{\textit{CBAtt}}: Class-based attention (CBAtt) \cite{sriker2022class} extracts features via ResNet50, and learns class-specific attention maps.  

\textbf{\textit{ConsultNet}}: ConsultNet \cite{9336317} uses a two-branch architecture based on DenseNet121, spatial and channel attention to learn discriminative features. 

\textbf{\textit{DuaLAnet}}: Dual lesion attention network (DuaLAnet) \cite{teixeira2020dualanet} consists of two asymmetric attention networks based on DenseNet169 and ResNet152.

\subsubsection{Region-guided methods}
\textbf{\textit{TSCN}}: Two-stream collaborative network (TSCN) \cite{CHEN2020221} creates a segmentation mask via U-Net and performs feature extraction on the masked region via DenseNet169.

\textbf{\textit{WSLM}}: Weakly supervised localization method (WSLM) \cite{JUNG202334} generates masks for pathology-containing regions and performs feature extraction via ResNet50. 

\textbf{\textit{RpSal}}: RpSal \cite{hermoza2020region} extracts features via a pyramid network followed by region proposal and saliency detection for simultaneous localization and classification. 

\textbf{\textit{LLAGnet}}: Lesion location attention guided network (LLAGnet) \cite{chen2019lesion} extracts features via DenseNet169, and uses weakly supervised attention to localize lesions in CXR images. 

\subsubsection{Semantic-guided methods}
\textbf{\textit{SEMM}}: SEMM \cite{yan2018weakly} extracts semantic features via DenseNet121 that are split into three branches using multi-map transfer learning. Features are concatenated across branches following class-wise pooling. 

\textbf{\textit{CheXGCN}}: Label co-occurrence learning framework based on graph convolution networks (CheXGCN) \cite{8961143} extracts features via DenseNet169, and captures co-occurrence relationships via a GCN. 

\textbf{\textit{SSGE}}: Semantic similarity graph embedding (SSGE) \cite{9430552} constructs a similarity graph from learned image features, and uses knowledge distillation to capture semantic similarities.

\textbf{\textit{TNELF}}: Triple network ensemble learning framework (TNELF) \cite{yang2023performance} uses ensemble learning based on DenseNet169, ResNet50, and EfficientNet-B4 backbones, and performs feature-wise concatenation for classification.

\subsection{Performance Evaluation}
Model performance was characterized via the Area Under the Curve (AUC) metric, which is commonly preferred for quantitative evaluation of multi-label CXR classification results. To do this, the area under the receiver operating characteristic (ROC) curve was first computed for each class. These areas were then averaged across classes. A higher AUC score indicates improved classification performance. Given a test set of CXR image $x_{q} = \left\{ x_{q1}, x_{q2}, ... x_{qN}\right\}$, the AUC of a classification model for the $c$th class is computed as:
\begin{align}
    AUC=\frac{\sum_{qi=1}^{N}\sum_{qj=1}^{N} \xi\left( {y}_{qi}^{c} < {y}_{qj}^{c} \right) \xi\left( \tilde{y}_{qi}^{c} < \tilde{y}_{qj}^{c} \right)}{\sum_{qi=1}^{N}\sum_{qj=1}^{N} \xi\left( {y}_{qi}^{c} < {y}_{qj}^{c} \right)}
    \label{eq:Eq7}
\end{align}
where $\xi(.)$ denotes an indicator function for the condition expressed via its input argument. The AUC metric reflects the discriminative ability of a classification model by quantifying the proportion of cases in which the ranking of predicted labels is aligned with the ranking of true labels.

\begin{table}[t!]
\centering
\captionsetup{justification=justified,width=\linewidth}
\caption{Performance of HydraViT and ablated variant models are listed as average$\pm$std AUC across labels. Results are shown for the subset of test samples where only a single label is present, the subset of test samples where only multiple co-occurring labels are present, and all test samples.}
\begin{center}
\resizebox{0.9\columnwidth}{!}
{%
\begin{tabular}{|c|c|c|c|c|c|}
\cline{1-6}  & w/o $f_{MBO}$ & w/o $f_{CE}$ & w/o $\tilde{y}_{i}^{A}$ & w/o init. & HydraViT \\ \hline
  Single & 77.0$\pm$8.5 & 76.8$\pm$8.8 & 79.7$\pm$8.8 & 78.0$\pm$8.5 & 79.8$\pm$8.6 \\ \cline{1-6} 
  Multiple     & 80.1$\pm$4.4 & 83.7$\pm$5.0 & 85.5$\pm$4.1 & 85.8$\pm$4.1 & 86.3$\pm$3.6 \\ \cline{1-6} 
  All & 79.0$\pm$6.0 & 80.7$\pm$6.3 & 83.3$\pm$5.9 & 82.8$\pm$6.0 & 83.8$\pm$5.8 \\ \hline
\end{tabular}%
}
\end{center}%
\label{tab:table1}
\end{table}

\begin{table*}[t]
\centering
\captionsetup{justification=justified,width=\linewidth}
\caption{Classification performance of competing methods on the ChestX-Ray14 dataset. Results are shown for attention-guided, region-guided, and semantic-guided baselines along with HydraViT. AUC is listed for each pathology label on separate rows; average AUC performance is also given as average$\pm$std across labels. Bold font marks the top performing method in each task.}
\begin{center}
\setlength\tabcolsep{2pt}
\renewcommand{\arraystretch}{1.4}
\resizebox{\textwidth}{!}
{%
\begin{tabular}{c|ccccc|cccc|cccc|c}
\cline{2-14}
                                    & \multicolumn{5}{c|}{Attention-guided methods}                                                                                                    & \multicolumn{4}{c|}{Region-guided methods}                                                    & \multicolumn{4}{c|}{Semantic-guided methods}                                                              &                                     \\ \cline{2-15} 
                                    & \multicolumn{1}{c|}{PCAN} & \multicolumn{1}{c|}{$\text{A}_{}^{3}$Net} & \multicolumn{1}{c|}{CBAtt}         & \multicolumn{1}{c|}{ConsultNet} & DuaLAnet & \multicolumn{1}{c|}{TSCN} & \multicolumn{1}{c|}{WSLM}          & \multicolumn{1}{c|}{RpSal} & LLAGnet & \multicolumn{1}{c|}{SEMM}          & \multicolumn{1}{c|}{CheXGCN} & \multicolumn{1}{c|}{SSGE}          & TNELF         & \multicolumn{1}{c|}{HydraViT}       \\ \hline
\multicolumn{1}{|c|}{Atelectasis}   & \multicolumn{1}{c|}{79.1} & \multicolumn{1}{c|}{77.9}                 & \multicolumn{1}{c|}{79.0}          & \multicolumn{1}{c|}{79.7}       & 78.3     & \multicolumn{1}{c|}{78.5} & \multicolumn{1}{c|}{79.0}          & \multicolumn{1}{c|}{77.5}  & 78.3    & \multicolumn{1}{c|}{79.2}          & \multicolumn{1}{c|}{78.6}    & \multicolumn{1}{c|}{79.2}          & 78.8          & \multicolumn{1}{c|}{\textbf{81.0}} \\ \hline
\multicolumn{1}{|c|}{Cardiomegaly}  & \multicolumn{1}{c|}{88.7} & \multicolumn{1}{c|}{89.5}                 & \multicolumn{1}{c|}{\textbf{91.0}} & \multicolumn{1}{c|}{90.9}       & 88.4     & \multicolumn{1}{c|}{88.7} & \multicolumn{1}{c|}{\textbf{91.0}} & \multicolumn{1}{c|}{88.1}  & 88.5    & \multicolumn{1}{c|}{88.1}          & \multicolumn{1}{c|}{89.3}    & \multicolumn{1}{c|}{89.2}          & 87.5          & \multicolumn{1}{c|}{90.4}          \\ \hline
\multicolumn{1}{|c|}{Consolidation} & \multicolumn{1}{c|}{75.9} & \multicolumn{1}{c|}{75.9}                 & \multicolumn{1}{c|}{76.0}          & \multicolumn{1}{c|}{77.9}       & 74.6     & \multicolumn{1}{c|}{75.4} & \multicolumn{1}{c|}{74.0}          & \multicolumn{1}{c|}{74.7}  & 75.4    & \multicolumn{1}{c|}{76.0}          & \multicolumn{1}{c|}{75.1}    & \multicolumn{1}{c|}{75.3}          & 75.6          & \multicolumn{1}{c|}{\textbf{82.2}} \\ \hline
\multicolumn{1}{|c|}{Edema}         & \multicolumn{1}{c|}{85.4} & \multicolumn{1}{c|}{85.5}                 & \multicolumn{1}{c|}{86.0}          & \multicolumn{1}{c|}{85.8}       & 84.1     & \multicolumn{1}{c|}{84.9} & \multicolumn{1}{c|}{86.0}          & \multicolumn{1}{c|}{84.6}  & 85.1    & \multicolumn{1}{c|}{84.8}          & \multicolumn{1}{c|}{85.0}    & \multicolumn{1}{c|}{84.8}          & 85.4          & \multicolumn{1}{c|}{\textbf{88.2}} \\ \hline
\multicolumn{1}{|c|}{Effusion}      & \multicolumn{1}{c|}{84.1} & \multicolumn{1}{c|}{83.6}                 & \multicolumn{1}{c|}{83.0}          & \multicolumn{1}{c|}{84.8}       & 83.2     & \multicolumn{1}{c|}{83.1} & \multicolumn{1}{c|}{84.0}          & \multicolumn{1}{c|}{83.1}  & 83.4    & \multicolumn{1}{c|}{84.1}          & \multicolumn{1}{c|}{83.2}    & \multicolumn{1}{c|}{84.0}          & 83.7          & \multicolumn{1}{c|}{\textbf{87.8}} \\ \hline
\multicolumn{1}{|c|}{Emphysema}     & \multicolumn{1}{c|}{94.4} & \multicolumn{1}{c|}{93.3}                 & \multicolumn{1}{c|}{93.0}          & \multicolumn{1}{c|}{92.9}       & 93.7     & \multicolumn{1}{c|}{93.0} & \multicolumn{1}{c|}{\textbf{95.0}} & \multicolumn{1}{c|}{93.6}  & 93.9    & \multicolumn{1}{c|}{92.2}          & \multicolumn{1}{c|}{94.4}    & \multicolumn{1}{c|}{94.8}          & 93.4          & \multicolumn{1}{c|}{90.8}          \\ \hline
\multicolumn{1}{|c|}{Fibrosis}      & \multicolumn{1}{c|}{81.9} & \multicolumn{1}{c|}{83.8}                 & \multicolumn{1}{c|}{82.0}          & \multicolumn{1}{c|}{83.4}       & 82.0     & \multicolumn{1}{c|}{83.3} & \multicolumn{1}{c|}{84.0}          & \multicolumn{1}{c|}{83.3}  & 83.2    & \multicolumn{1}{c|}{83.3}          & \multicolumn{1}{c|}{83.4}    & \multicolumn{1}{c|}{82.7}          & \textbf{84.9} & \multicolumn{1}{c|}{82.4}          \\ \hline
\multicolumn{1}{|c|}{Hernia}        & \multicolumn{1}{c|}{92.8} & \multicolumn{1}{c|}{93.8}                 & \multicolumn{1}{c|}{93.0}          & \multicolumn{1}{c|}{92.8}       & 89.5     & \multicolumn{1}{c|}{92.1} & \multicolumn{1}{c|}{88.0}          & \multicolumn{1}{c|}{91.7}  & 91.6    & \multicolumn{1}{c|}{93.4}          & \multicolumn{1}{c|}{92.9}    & \multicolumn{1}{c|}{93.2}          & \textbf{94.4} & \multicolumn{1}{c|}{88.7}          \\ \hline
\multicolumn{1}{|c|}{Infiltration}  & \multicolumn{1}{c|}{71.1} & \multicolumn{1}{c|}{71.0}                 & \multicolumn{1}{c|}{72.0}          & \multicolumn{1}{c|}{70.9}       & 70.8     & \multicolumn{1}{c|}{70.3} & \multicolumn{1}{c|}{71.0}          & \multicolumn{1}{c|}{69.5}  & 70.3    & \multicolumn{1}{c|}{71.0}          & \multicolumn{1}{c|}{69.9}    & \multicolumn{1}{c|}{71.4}          & \textbf{72.2} & \multicolumn{1}{c|}{71.2}          \\ \hline
\multicolumn{1}{|c|}{Mass}          & \multicolumn{1}{c|}{83.9} & \multicolumn{1}{c|}{83.4}                 & \multicolumn{1}{c|}{83.0}          & \multicolumn{1}{c|}{84.8}       & 83.7     & \multicolumn{1}{c|}{83.3} & \multicolumn{1}{c|}{85.0}          & \multicolumn{1}{c|}{82.6}  & 84.1    & \multicolumn{1}{c|}{84.7}          & \multicolumn{1}{c|}{84.0}    & \multicolumn{1}{c|}{84.8}          & 84.5          & \multicolumn{1}{c|}{\textbf{87.4}} \\ \hline
\multicolumn{1}{|c|}{Nodule}        & \multicolumn{1}{c|}{80.9} & \multicolumn{1}{c|}{77.7}                 & \multicolumn{1}{c|}{79.0}          & \multicolumn{1}{c|}{78.9}       & 80.0     & \multicolumn{1}{c|}{79.8} & \multicolumn{1}{c|}{81.0}          & \multicolumn{1}{c|}{78.9}  & 79.0    & \multicolumn{1}{c|}{81.1}          & \multicolumn{1}{c|}{80.0}    & \multicolumn{1}{c|}{\textbf{81.2}} & 80.0          & \multicolumn{1}{c|}{78.3}          \\ \hline
\multicolumn{1}{|c|}{P-Thickening}  & \multicolumn{1}{c|}{80.6} & \multicolumn{1}{c|}{79.1}                 & \multicolumn{1}{c|}{78.0}          & \multicolumn{1}{c|}{79.6}       & 79.6     & \multicolumn{1}{c|}{78.2} & \multicolumn{1}{c|}{80.0}          & \multicolumn{1}{c|}{79.3}  & 79.8    & \multicolumn{1}{c|}{\textbf{80.8}} & \multicolumn{1}{c|}{79.5}    & \multicolumn{1}{c|}{79.5}          & 78.4          & \multicolumn{1}{c|}{79.7}          \\ \hline
\multicolumn{1}{|c|}{Pneumonia}     & \multicolumn{1}{c|}{74.6} & \multicolumn{1}{c|}{73.7}                 & \multicolumn{1}{c|}{73.0}          & \multicolumn{1}{c|}{74.0}       & 73.5     & \multicolumn{1}{c|}{73.1} & \multicolumn{1}{c|}{73.0}          & \multicolumn{1}{c|}{74.1}  & 72.9    & \multicolumn{1}{c|}{74.0}          & \multicolumn{1}{c|}{73.9}    & \multicolumn{1}{c|}{73.3}          & 73.4          & \multicolumn{1}{c|}{\textbf{75.9}} \\ \hline
\multicolumn{1}{|c|}{Pneumothorax}  & \multicolumn{1}{c|}{88.1} & \multicolumn{1}{c|}{87.8}                 & \multicolumn{1}{c|}{87.0}          & \multicolumn{1}{c|}{87.4}       & 86.6     & \multicolumn{1}{c|}{88.1} & \multicolumn{1}{c|}{89.0}          & \multicolumn{1}{c|}{87.9}  & 87.7    & \multicolumn{1}{c|}{87.6}          & \multicolumn{1}{c|}{87.6}    & \multicolumn{1}{c|}{88.5}          & 86.2          & \multicolumn{1}{c|}{\textbf{89.4}} \\ \hline
\hline
\multicolumn{1}{|c|}{Average}          & \multicolumn{1}{c|}{83.0$\pm$6.4} & \multicolumn{1}{c|}{82.6$\pm$6.8}                 & \multicolumn{1}{c|}{82.3$\pm$6.6}          & \multicolumn{1}{c|}{83.1$\pm$6.5}       & 82.0$\pm$6.2    & \multicolumn{1}{c|}{82.3$\pm$6.6} & \multicolumn{1}{c|}{83.0$\pm$6.7}          & \multicolumn{1}{c|}{82.1$\pm$6.7}  & 82.4$\pm$6.6    & \multicolumn{1}{c|}{83.0$\pm$6.2}          & \multicolumn{1}{c|}{82.6$\pm$6.8}    & \multicolumn{1}{c|}{83.0$\pm$6.7}          & 82.7$\pm$6.5         & \multicolumn{1}{c|}{\textbf{83.8}$\pm$\textbf{5.8}} \\ \hline
\end{tabular}%
}
\end{center}%
\label{tab:table2}
\end{table*}

\section{Results}

\subsection{Ablation Studies}
Several ablation studies were conducted to demonstrate the contribution of the individual components in HydraViT to method performance. First, we examined the effect of the spatial encoder, the context encoder, and the multi-branch output modules. For this purpose, HydraViT that comprises all three modules was compared against a variant that contained only the SE module $f_{SE}$ that was augmented with a softmax output layer for multi-label classification, a variant that contained the SE and CE modules $f_{SE}+f_{CE}$ augmented with a softmax output layer, and a variant that contained the SE and MBO modules $f_{SE}+f_{MBO}$. Figure \ref{fig:fig2} displays ROC curves and respective AUC metrics for the compared models, separately for each pathology label and on average across labels. On average across labels, HybdraViT outperforms $f_{SE}$ by 5.9\%, $f_{SE}+f_{CE}$ by 4.8\%, and $f_{SE}+f_{MBO}$ by 3.1\% AUC. We also find that $f_{SE}+f_{MBO}$ consistently outperforms $f_{SE}$ and that HydraViT consistently outperforms $f_{SE}+f_{CE}$ across labels. For these cases, the most notable improvements due to the introduction of the MBO module are observed for relatively rare labels such as `Cardiomegaly', `Infiltration', and `Pneumonia' that often co-occur with other pathology. This finding indicates the importance of the MBO module over a conventional softmax classification layer in multi-label classification. We also observe that HydraViT consistently outperforms $f_{SE}+f_{MBO}$ across labels, with more notable improvements for labels such as `Nodule', `Mass', `Fibrosis', and `Atelectasis' where pathology can manifest with an atypical intensity distribution at both local and global scale. This result indicates the importance of the self-attention mechanism in CE to capture local and global contextual features of CXR images.  

Next, we examined the benefits of the multi-task training of the MBO module in HydraViT for multi-label classification. HydraViT was compared against an aggregated variant that contained the SE and CE modules $f_{SE}+f_{CE}$ augmented with a softmax output layer based on an aggregated output vector, and an ensemble variant that utilized multiple $f_{SE}+f_{CE}$ models with uni-dimensional output variables trained separately for each individual label. Figure \ref{fig:fig3} illustrates classification performance for the compared models. Among the variants, either the aggregated or the ensemble variant yields better performance in some labels, and the two variants perform similarly in remaining labels. That said, HydraViT consistently outperforms the two variants across all labels. On average, HydraViT outperforms the aggregated variant by 4.8\% suggesting that multi-task training based on separate output branches helps improve sensitivity to individual labels. HydraViT also outperforms the ensemble variant by 9.4\% suggesting that the aggregated output branch in HydraViT helps improve capture of co-occurrence relationships among pathology. 

We finally assessed the benefits of the MBO module, CE module, aggregated output vector and MBO weight initialization in HydraViT on classification performance for single versus multiple labels. A variant that excluded the MBO module from HydraViT (w/o $f_{MBO}$), a variant that excluded the CE module (w/o $f_{CE}$), a variant that excluded the aggregated output vector and the associated multi-label cross-entropy and consistency loss terms (w/o $\tilde{y}{i}^{A}$), and a variant with zero initialization of weights in the MBO module (w/o $init.$) were considered. Table \ref{tab:table1} lists the classification performance of the compared models on the subset of test samples that contain only a single label, on the subset of the test sample that contains multiple co-occurring labels, and on the entire test set. We find that HydraViT outperforms the variants in all cases, albeit performance benefits are more notable for the multiple label case. In the single-label case, HydraViT outperforms w/o $f_{MBO}$ by 2.8\%, w/o $f_{CE}$ by 3.0\%, w/o $\tilde{y}{i}^{A}$ by 0.1\% and w/o $init.$ by 1.8\% AUC. In the multiple-label case, HydraViT outperforms w/o $f_{MBO}$ by 6.2\%, w/o $f_{CE}$ by 2.6\%, w/o $\tilde{y}{i}^{A}$ by 0.8\% and w/o $init.$ by 0.5\%. Across the test set, HydraViT outperforms w/o $f_{MBO}$ by 4.8\%, w/o $f_{CE}$ by 3.1\%, w/o $\tilde{y}{i}^{A}$ by 0.5\% and w/o $init.$ by 1.0\%.
Note also that HydraViT yields the lowest standard deviation in AUC across labels, indicating an improvement in homogeneity of classification performance across distinct pathology.

\subsection{Comparison Studies}
We comparatively demonstrated the performance of HydraViT in multi-label CXR classification against several state-of-the-art methods including attention-guided (PCAN \cite{ZHU2022102137}, $\text{A}_{}^{3}$Net \cite{WANG2021101846}, CBAtt \cite{sriker2022class }, ConsultNet \cite{9336317}, DuaLAnet \cite{teixeira2020dualanet}), region-guided (TSCN \cite{CHEN2020221}, WSLM \cite{JUNG202334}, RpSal \cite{hermoza2020region}, LLAGnet \cite{ chen2019lesion}), and semantic guided models (SEMM \cite{yan2018weakly}, CheXGCN \cite{8961143}, SSGE \cite{9430552}, TNELF \cite{yang2023performance}). Table \ref{tab:table2} lists the classification performance of the competing models separately for each pathology label, and on average across labels. In terms of pathology labels, the performance improvements offered by HydraViT are most notable for `Atelectasis', `Consolidation', `Edema', `Effusion', `Mass', `Pneumonia', `Pneumothorax' that can show a relatively broad spatial distribution across CXR images. HydraViT yields comparable performance to most baselines for labels such as `P-Thicknening', `Infiltration', `Fibrosis', and `Cardiomegaly'. Meanwhile, several baselines can yield higher performance than HydraViT for `Emphysema', `Hernia', `Nodule'. On average, HydraViT outperforms attention-guided baselines by 1.2\%, region-guided baselines by 1.4\%, and semantic-guided baselines by 1.0\% AUC. Furthermore, HydraViT also achieves the lowest standard deviation in AUC across labels among competing methods, indicating an improvement in homogeneity of classification performance across distinct pathology. These findings suggest that HydraViT improves performance and reliability in multi-label CXR classification.

\begin{figure*}[t]
\centering
\includegraphics[width=0.9\linewidth]{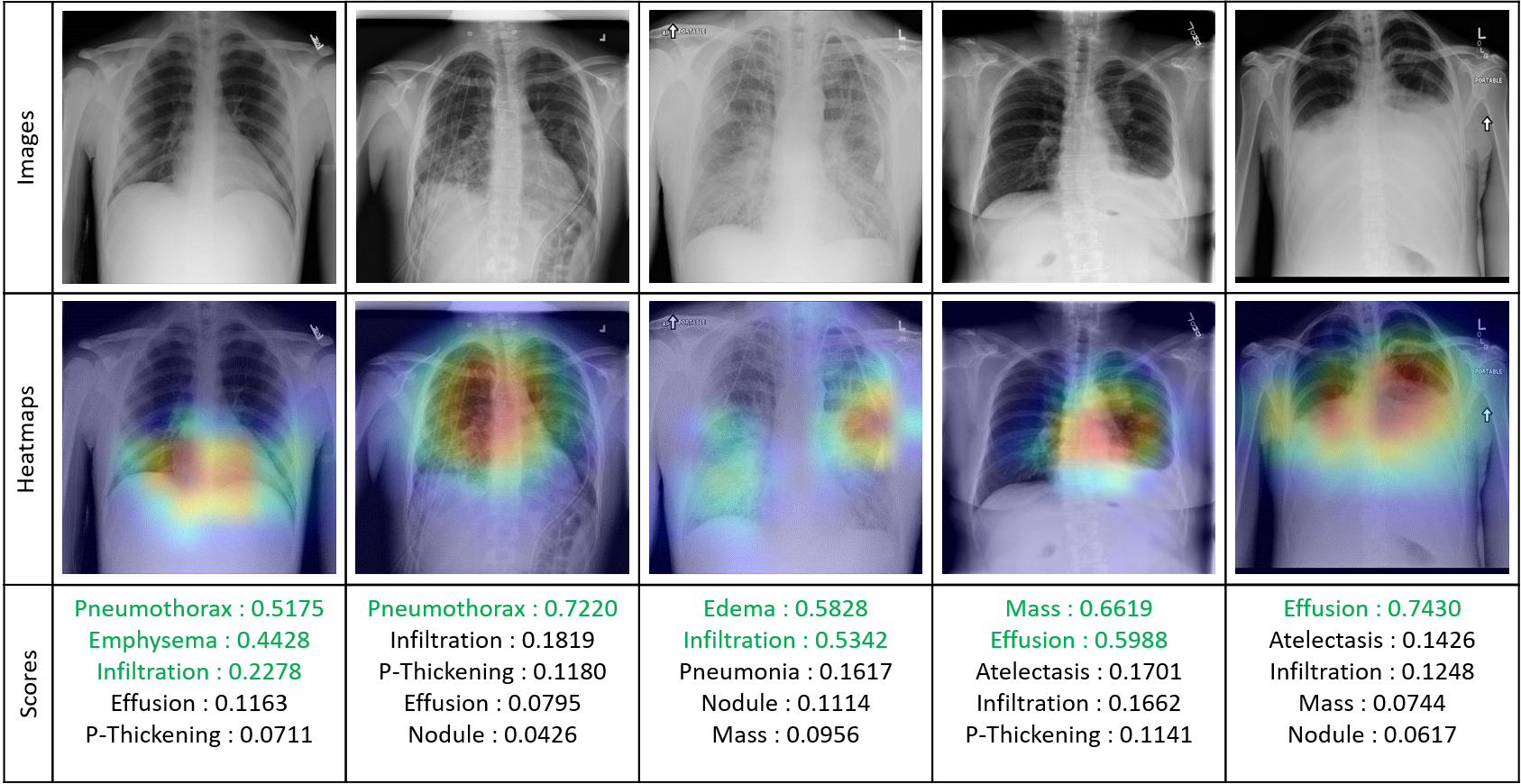}
\caption{Representative CXR images from the ChestX-Ray14 dataset, and respective multi-label predictions generated by HydraViT. The top-5 predicted pathologies and probability scores are listed, and the ground truth labels are marked in green. GradCAM-derived heatmaps for the CXR images are also given to highlight pathological regions.}
\label{fig:fig4}
\end{figure*}

Finally, we examined the performance of HydraViT in multi-label classification qualitatively by inspecting the predicted labels in representative CXR images. Figure \ref{fig:fig4} displays the CXR images, corresponding heatmaps extracted via the GradCAM method \cite{8237336} that highlight salient regions relevant to pathology, and the scores for top-5 predicted labels. The ground-truth labels are annotated in green font. We observe that the top-ranked labels by HydraViT are closely aligned with the ground-truth labels. In each case, the average score estimated by HydraViT for the ground-truth labels is significantly higher than the average score for non-present labels within the list of top-5 (e.g., 0.4 versus 0.1, 0.7 versus 0.1, 0.6 versus 0.1, 0.6 versus 0.2 for the representative samples presented in Figure \ref{fig:fig4}). These results indicate that HydraViT yields an accurate estimation of pathology in multi-label CXR classification.

\section{Discussion and Conclusion}
In this study, we proposed a novel deep learning method to improve performance in multi-label CXR classification of thoracic diseases. The proposed HydraViT model uses a hybrid convolutional-transformer backbone to extract contextualized embeddings of CXR images, and a multi-branch output module with adaptive weights to improve capture of co-occurring pathology. While branched output modules have been previously considered for multi-task learning problems in the machine learning literature, to our knowledge, HydraViT is the first method to devise a multi-branch architecture that comprises output heads for individual and aggregated labels in multi-label classification. 

A set of ablation studies were conducted to demonstrate the contribution of individual design elements in HydraViT. These studies indicate that the introduction of the transformer-based context encoder helps significantly boost classification performance. They also indicate that the multi-branch output module in HydraViT yields elevated performance over training separate network models for each label and training a single model with only separate heads for individual labels. Note that there is also a computational benefit for training a single multi-branch architecture as in HydraViT, which is nearly 14 times faster compared to sequential training of single-branch architectures for each pathology label separately. 

HydraViT was comparatively demonstrated against state-of-the-art deep learning methods for multi-label CXR classification. Attention-guided, region-guided, and semantic-guided baselines were considered. While there were occasional cases where a competing baseline yielded comparable or higher scores for 1-3 pathology labels out of 14, HydraViT generally outperformed baselines in the majority of labels. Across all labels, HydraViT yielded the highest average performance in multi-label classification. The success of HydraViT in identifying the presence of multiple pathologies was also corroborated via visual inspections. Therefore, our results suggest that HydraViT is a promising approach for CXR-based classification of pathology in thoracic diseases.

Several technical limitations can be addressed in future work to further improve the performance of HydraViT. A hybrid convolutional-transformer architecture was adopted here to maintain computational efficiency while extracting contextual embeddings. Sensitivity for long-range context can be further boosted by adopting a pure transformer architecture at the expense of elevated model complexity. In those cases, low-rank approximations on the self-attention matrix or pyramidal transformer architectures with limited attention windows can help improve efficiency \cite{bedel2023bolt}. Note that the pathology annotations for the CXR dataset analyzed in this study were mined from radiological reports via language models, so they contain an inherent level of noise \cite{doi:10.1148/radiol.2019191293}. Such noisy levels can introduce a degree of bias in trained models that take annotations as ground truth. Learning procedures that take into account the possibility of erroneous labels can help boost classification performance. While the loss terms used here included learnable weights to maintain the balance between separate labels, classification models can still be susceptible to class imbalance. Discrimination among labels can instead be performed based on embedding distances in a latent space for CXR images to alleviate such biases \cite{9810181,OZTURK2023118938}. Here, a context encoder module equipped with self-attention mechanisms was used to enable the model to focus on pathological regions in CXR images. For improved localization, anomaly detection on CXR images based on generative approaches such as diffusion models could be utilized \cite{dar2022adaptive,ozbey2022unsupervised}. Finally, sensitivity for co-occurrence relationships can be enhanced by adopting focal modulation networks instead of transformers with self-attention filtering \cite{yang2022focal}.

\bibliographystyle{IEEETran} 
\bibliography{IEEEabrv,HydraViT}

\end{document}